\documentclass[english,aps,prc, twocolumn,nofootinbib,0superscriptaddress,longbibliography]{revtex4-1}

\usepackage{epsfig}
\usepackage{graphicx}
\usepackage{amsmath}
\usepackage{lipsum}
\usepackage{color}


\def\orcid#1{\kern .08em\href{https://orcid.org/#1}{\includegraphics[keepaspectratio,width=0.7em]{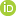}}}

\usepackage{babel}
\usepackage[colorlinks,linkcolor=blue,anchorcolor=blue,citecolor=blue,urlcolor=blue,filecolor=black]{hyperref}

\newcommand{\be}{\begin{equation}}
\newcommand{\ee}{\end{equation}}
\newcommand{\ba}{\begin{eqnarray}}
\newcommand{\ea}{\end{eqnarray}}

\begin{document}
\title{How to extract the electromagnetic response of $^6$He in relativistic collisions}
\author{C.A. Bertulani}
\email[]{Email: carlos.bertulani@tamuc.edu}
\affiliation{Department of Physics and Astronomy, Texas A\&M University-Commerce,  Texas 75429-3011, USA}
\affiliation{Institut f\"ur Kernphysik,  Technische Universit\"at Darmstadt, 64289 Darmstadt, Germany}
\affiliation{Helmholtz Research Academy Hesse for FAIR, D–64289 Darmstadt, Germany}
\date{\today}

\begin{abstract}
I investigate the difficulties in obtaining the electromagnetic response of light, halo-like, nuclei using reactions at radioactive beam facilities. A relativistic coupled-channels theory for the calculation of dissociation cross sections of halo nuclei is compared to first-order perturbation theory. A comparison with semiclassical models frequently used in experimental analysis is also performed. It is shown that the effects of relativity and of the nuclear interaction lead to sizable effects in the extraction of the electromagnetic response of $^6$He projectiles.
\end{abstract}

\maketitle

%\tableofcontents

\section{Introduction}
The investigation of nuclear reactions induced by unstable nuclei is important for both nuclear physics and nuclear astrophysics \cite{AUMANN2020103753}.
Measurements of nuclear reactions involving unstable nuclei as targets is not feasible and the most viable method to experimentally study their nuclear properties is to resort to  indirect methods using the unstable nuclei as projectiles \cite{Aumann:05}. Among the methods that have been used for this purpose in radioactive beam facilities, I cite the Coulomb dissociation of radioactive projectiles. The pioneer work in Ref. \cite{BAUR1986188} showed that this method can be used to extract radiative capture cross sections of the type (n,$\gamma$), (p,$\gamma$), and ($\alpha, \gamma$) of interest for nuclear astrophysics using detailed balance. Subsequent experiments have demonstrated the usefulness of the Coulomb dissociation method for this purpose (see, e.g., Refs. \cite{PhysRevLett.67.808,motobayashi1991determination,MotobayashiPRL73.2680,Horvath_2002,PhysRevC.88.065808,PhysRevC.63.065806,PhysRevLett.90.232501,Marganiec_2016}) as well as for various other studies of nuclear structure and nuclear astrophysics such as the electromagnetic response of radioactive nuclei and the excitation of pygmy resonances and its connection to the physics of neutron stars
(see, e.g., Refs \cite{Kobayashi:89,Sackett:93,IekiPRL70.730,Zinser1995,SHIMOURA199529,NAKAMURA2003C301,wieland:2009:PRL,rossi:2013:PRL,savran:2013:PPNP}).

Coulomb breakup reactions are carried out at radioactive beam facilities with projectile bombarding energies of a few hundred MeV/nucleon impinging on stable targets $A$ such as ${^9}$Be, $^{12}$C, or $^{208}$Pb. It is the purpose of this article to report a study of various reaction mechanisms and to guide experimental analysis for  a proper extraction of the electromagnetic response of radioactive nuclei. I will discuss reactions of the type $a + A \rightarrow  c  + N_1 +N_2 +A$, i.e., when the projectile is dissociated into a core nucleus, $c$, and two nucleons, $N_i$ ($i=1,2$).  In particular, I will apply the results to  $^6$He dissociation, which has been studied experimentally  \cite{PhysRevC.59.1252}. Recently, highly accurate data were obtained and experimental analysis is in progress \cite{Lehr}. 

Besides dissociation via the Coulomb interaction, the nuclear interaction also contributes appreciably to projectile dissociation. That poses a hurdle to the experimental analysis because the strong interaction between nuclei is not as well known as its Coulomb counterpart. It has been a challenge for theorists and experimentalists alike during many decades to determine which interaction is dominant when distinct kinematic conditions are selected.   
Additionally, the modeling of direct reactions with relativistic projectiles  involves formidable challenges: (1) Because reactions carried out at few hundred MeV/nucleon involve a sizable increase of the projectile rest mass, a relativistic covariant reaction theory is desirable. (2) An optical model potential for direct reactions with relativistic nuclei is not considered properly in most theoretical studies published so far. (3) The inclusion of couplings to, from, and within the nuclear structure continuum is important for reactions involving weakly bound nuclei. 
Various theoretical publications have addressed items 1 and 2, for example Refs. \cite{aleixo:1989:NPA,BertulaniPRC49.2839,ESBENSEN1995107,PhysRevC.58.2864,PhysRevLett.95.082502,BertulaniPRL94.072701,PhysRevC.83.024907,Yabana2008,OgataPTP.123.701,doi:10.7566/JPSCP.32.010028,Hebborn_2023,PhysRevC.96.054607} and many others. Item 3 has been considered in many publications, although the calculations have mostly been performed ignoring the effects of items 1 and 2 (see, e.g., \cite{PhysRevLett.95.082502,CHATTERJEE2000477,Hebborn2018,Bertulani2020,PhysRevC.107.014607,SHUBHCHINTAK201499,PhysRevC.79.024607,Casal_2020,HAGINO2022103951,DRISCHLER2021136777,PhysRevC.103.014604}).

In this article I revisit the theoretical challenges in describing breakup reactions  at few hundred MeV/nucleon. I include new theoretical developments. My goal is to show how one can extract the electromagnetic response of halo nuclei. I apply my theory to the dissociation of $^6$He projectiles incident on carbon, tin and lead targets. I discuss the challenges encountered to develop a covariant theory for breakup reactions at relativistic energies and propose additional methods to tackle the problem.  Medium corrections are included by the construction of optical potentials using effective nucleon-nucleon interactions and nucleon-nucleon scattering observables as building blocks. This last assumption is justified because the collision is sudden and nucleons tend to be involved individually in a frozen configuration during the reaction. The most relevant part of the optical potential in high energy collisions is the imaginary part which emerges from absorption due to binary nuclear collisions. The increase of the rest mass of the nuclei and the effects of relativistic dynamics on the optical and Coulomb potentials is studied using minimal and transparent approximations that will help to clarify the main challenges of the reaction mechanism.

In Secs. \ref{cceqs}-\ref{rnd} I discuss the modifications in the dynamic equations for the breakup process to include relativistic corrections and couplings with the continuum. 
In Sec. \ref{HET} I make use of a simplified three-body model to test the effects of reaction dynamics on the breakup of $^6$He. Numerical results for the dissociation of $^6$He are shown in Sec. \ref{bu6He}. My conclusions are presented in Sec. \ref{concl}.
Whenever possible, I will use a simplified notation to focus on where improvements could be done in future works.

%\section{Coupled-channels equations }
\section{ Excitation amplitudes\label{cceqs}}
I adopt the relativistic eikonal CDCC (continuum discretized coupled-channels)  method proposed in Ref. \cite{BertulaniPRL94.072701}. The coupled channels equation reads
\begin{equation}
{i\hbar v}\dfrac{d}{d z}{\cal S}_c({\bf b},z) = \sum_{c'}\left\langle
\Phi_{c}|{\cal H}_{int}({\bf b},z)|\Phi_{c'}\right\rangle
{\cal S}_{c'}({\bf b},z)\ e^{i {E_{cc'}  z\over \hbar v}}, \label{cceq4}
\end{equation}
where ${\cal S}_c({\bf b},z)$ is the reaction {\it S}-matrix for a collision at the impact parameter ${\bf b}$, $v$ is the (supposedly) undisturbed projectile velocity, a good approximation for high-energy collisions, ${\cal H}_{int}$ is the interaction hamiltonian, and $c$ are the channel indices \{$i$, $\ell$, $m$\}. The index $i>0$
($i=0$) denotes the $i$th discretized-continuum (ground) state, and $\ell$ and $m$\ are, respectively, the orbital angular momentum
and its projection along the $z$-axis taken to be parallel to the incident beam. $E_{cc'}=E_{c'}-E_c$ is the excitation energy, and $\Phi_c$ are the internal wave functions of the projectile.  By solving these equations with the initial condition,
$
{\cal S}_c(b,-\infty) = \delta_{c0}, $
one can obtain the probability that a channel $c$ is populated in the reaction, namely, $|{\cal S}_{c}(b,\infty)|^2$.

Equations \eqref{cceq4} were obtained from the Klein-Gordon equation  applied to the wave function of the projectile,  neglecting terms of the type $\nabla^2 {\cal S}_{c'}({\bf b},z)$ compared to $ik\partial  {\cal S}_{c'}({\bf b},z)/\partial z$, where $k^2= E^2-Mc^2$, with $E$ being the total energy of the projectile including its rest mass $M$. Equation \eqref{cceq4} is Lorentz invariant if the interaction Hamiltonian ${\cal H}_{int}$ is a scalar potential $U$ transforming as the time-like component of a four-vector. The scattering amplitude for the transition from the ground state to the continuum $0\rightarrow c$, including higher-order $c'c''$ couplings, is given by
\begin{equation}
f_c({\bf q}) = -{ik\over 2\pi} \int d{\bf b} \exp[i{\bf q.b}]\left[ {\cal S}_{c}({\bf b},\infty)-\delta_{0c}\right], \label{amp}
\end{equation}
where ${\bf  q} = {\bf k'} - {\bf k}$ is the momentum transfer in the reaction. Most cases of interest involve momentum transfers much smaller than the momentum of the impinging projectile. Thus, one can use the expression valid for elastic scattering $q=2k\sin (\theta/2)$, where $\theta$ is the scattering angle. 

The equations above include transitions in the continuum if a discretization procedure in the continuum is adopted. 
In my calculations, a continuum wave function in channel $c$ with energy $E_c$ can be discretized by using the simple bin discretizaton method \cite{BERTULANI1992163}
\begin{equation}
\left| E_c\right> = \int dE'_c \, \Gamma (E'_c) \left| E'_c\right>,
\label{gamm1}
\end{equation}
where $\Gamma (E'_c)$ is an appropriately ortho-normalized function peaked at energy $E_c$ and with a width $\Delta E_c$. I employ the simplest discretization method 
\begin{equation}
\Gamma(E_j) = \left\{ \begin{matrix}
{1\over \Delta E }     & {\rm if } \ \ \ (j-1)\Delta E < E_c < j \Delta E ,   \\
& \\
  0.    &  \ \ \ {\rm otherwise.}
\end{matrix} \right.
\label{gamm2}
\end{equation}
The inelastic cross section is obtained as
${d\sigma_c/ d\Omega} = |f_c({\bf q}, E_c)|^2 .$
For simplicity, the angular momentum and other quantum numbers are not displayed explicitly.
This formalism is known as relativistic continuum discretized coupled channels, with acronym RCDCC  \cite{BertulaniPRL94.072701}. It is an improvement over the non-relativistic coupled channels procedures, with a better description of experiments for bombarding energies around and above 100 MeV/nucleon.  

To include absorption at small impact parameters, the {\it S}-matrix ${\cal S}_c({\bf b},\infty)$ obtained by solving Eq. \eqref{amp} is corrected by the modification
\begin{eqnarray}
{\cal S}_c({\bf b},\infty) &\rightarrow& {\cal S}_c({\bf b},\infty) \exp\{i\chi(b)\}, \nonumber \\
\chi(b) &=&  \int {\rho}_{P}(q){\Gamma}(q){\rho}_{T}(q) J_0(qb)q \,dq ,\label{eik7}
\end{eqnarray} 
where $J_{0}$ is the ordinary Bessel function of zeroth-order, and the nucleon-nucleon scattering profile function is parametrized as \cite{Ray:1979}
\begin{equation}
{\Gamma}(q)=\frac{i}{4\pi }\sigma_{NN}e^{-\beta_{NN}q^2} .\label{eik8}
\end{equation}
In the equation above, $\sigma_{NN}$ is the total nucleon-nucleon cross section, and $\beta_{NN}$ is the momentum dependence parameter. $\rho_P$ ($\rho_T$) is the Fourier transform of the ground state density of the projectile (target). Tables with the energy dependence of these parameters are found in Refs. \cite{HUSSEIN1991279,AUMANN2020103753}. I add to the imaginary part of Eq. \eqref{eik7}  corrections due to Coulomb scattering at large impact parameters. They are properly accounted for by using the simple expression $\chi\rightarrow \chi+ \chi_c$ with $\chi_c(b) = 2\eta \ln (kb)$, where  $\eta = Z_PZ_T e^2/\hbar v$ is the Sommerfeld parameter \cite{BERTULANI1993158}. 

The coupled channels method described  above can be used with any nuclear structure model, either a two-body, three-body, or many-body model  to calculate the matrix elements $\left\langle \Phi_{c}|{\cal H}_{int}(b,z)|\Phi_{c'}\right\rangle$ in Eq. \eqref{cceq4}. First order excitation amplitudes can be obtained by using ${\cal S}_c(b,z)=\delta_{c0}$ on the right-hand side of Eq. \eqref{cceq4}, yielding
\begin{eqnarray}
{\cal S}_c(b,\infty) &=& -{i\over \hbar v}\int_{-\infty}^\infty dz \left\langle
\Phi_{c}|{\cal H}_{int}(b,z)|\Phi_0\right\rangle
e^{i{E_{0c}  z \over \hbar v}}\nonumber \\
&\times &  \exp\{i\chi(b)\}. \label{cceq9}
\end{eqnarray}

In the next sections, I describe how the RDCC equations reduce to results obtained with eikonal scattering waves and first-order perturbation theory. It is also worthwhile to compare them to semiclassical methods  extensively used in the analysis of nucleus-nucleus inelastic scattering. One major task is the treatment of relativistic corrections in the interaction Hamiltonian ${\cal H}_{int}$. While this can be achieved straightforwardly  in the case of the Coulomb interaction, in the nuclear case it can only be done in an approximate way by mimicking the Lorentz transformations of the Coulomb field \cite{OgataPTP.123.701}. 

 \section{Quantum and semiclassical angular distributions \label{QScat}}
 
 \subsection{Exact scattering amplitudes}
I will assume that the nuclei possess spherical symmetry, leading to a simplification of the integration in Eq. \eqref{amp}.  The angular distribution of the inelastically scattered particles is obtained from the 
amplitudes ${\cal S}_c(b,\infty)$. The excitation of the channel state $c\neq 0$ becomes 
\begin{equation}
f_{c}^\mu(\theta)=ik\int_0^\infty db b J_\mu(qb){\cal S}_c(b, \infty). \label{cceq6}
\end{equation}
The index $\mu = M_c - M_0$ denotes the change of the magnetic quantum number $M_i$ associated with the total angular momentum $J_i$ of the excited nucleus. The inelastic scattering cross section is obtained by an average over the initial spin and a sum over the final spins:
\begin{equation}
{d\sigma_{c} \over d \Omega} = {1\over 2J_0+1} \sum_{M_0,M_{c}}\left| f_{c}^\mu(\theta, E_c) \right|^2. \label{cceq7}
\end{equation}

For large bombarding energies  $q= 2k\sin \theta/2\simeq k\theta$, and $d\Omega=2\pi q dq/k^2$, leading to 
\begin{eqnarray}
\sigma &=& {2\pi\over 2J_0+1} \sum_{M_0,M_{c}}\int db\, b \int db' \, b'  \int dq \, q J_\mu(qb) J_\mu(qb')\nonumber \\
&\times& {\cal S} (b,\infty) {\cal S}^*(b',\infty) \nonumber \\
&=& {1\over 2J_0+1} \int db \, b \left|{\cal S}_c(b,\infty)\right|^2, \label{cceq12}
\end{eqnarray}
where, in the last step, the completeness relation of the Bessel functions, 
\begin{equation}
\int dq \, qJ_\mu(qb)J_\mu(qb')={1\over b}\delta(b-b') ,\label{cceq14}
\end{equation}
was used.
Equations (\ref{cceq6}-\ref{cceq12}) describe the angular distributions and total cross sections with the inclusion of any desired number of channel couplings after solving Eqs. \eqref{cceq4} for $ {\cal S} (b,\infty)$.

\subsection{First-order perturbation theory}
In first-order perturbation theory, one inserts ${\cal S}_{c'}({\bf b},\infty) = \delta_{c0}$ on the right-hand side of  Eq. \eqref{cceq4}.  As shown in Ref. \cite{BERTULANI1993158} [their Eqs. (11)-(15)],  for large impact parameters  the amplitudes ${\cal S}_c(b, \infty)$ can be calculated analytically and Eqs. (\ref{cceq6}) and (\ref{cceq7}) yield the same results as those already obtained in Ref.  \cite{BERTULANI1988299,BERTULANI1993158} where the differential Coulomb excitation cross sections are given by
\begin{equation}
{d^2\sigma \over d\Omega dE} = {1\over E} \sum_{\pi L} {dn_{\pi L} \over d\Omega }(E,\theta)  \sigma_\gamma^{\pi L} (E) \label{d2sfirst}
\end{equation}
and
\begin{equation}
{d\sigma \over  dE} = {1\over E} \sum_{\pi L} n_{\pi L} (E) \sigma_\gamma^{\pi L} (E). \label{dsCoulde}
\end{equation}
The photonuclear cross section for  multipolarity $\pi L$ is related to the response functions as \cite{BERTULANI1988299},
\begin{equation}
  \sigma_\gamma^{\pi L} (E) = {(2\pi)^3 (L+1) \over L [(2L+1)!!]^2} \left( {E\over \hbar c}\right)^{2L-1} {dB(\pi L, E) \over dE}.
  \end{equation}
  The equivalent photon  numbers are given by \cite{BERTULANI1988299}
 \begin{eqnarray} 
{dn_{\pi L} \over d\Omega }(E,\theta) &=& Z^2_T \alpha \left({E k \over \gamma \hbar v}\right)^2 {L [(2L+1)!!]^2 \over (2\pi)^3 (L+1)} \nonumber \\
&\times& \sum_M |G_{\pi LM} (c/v) |^2 |\Omega_M (q)|^2, \label{virtpiL}
\end{eqnarray}
where the functions $G_{\pi \lambda m}(c/v)$ are defined in Ref. \cite{WINTHER1979518}.
The function $\Omega_M(q)$ is given by \cite{BERTULANI1993158} 
\begin{equation}
\Omega_M(q) = \int_0^\infty db \, b J_M(qb) K_M\left( {Eb\over \gamma \hbar v}\right) \exp[i\chi(b)], \label{Omega}
  \end{equation}
where $J_M$  and $K_M$ are the Bessel and modified Bessel functions.  The virtual photon numbers entering Eqs. \eqref{dsCoulde} are
\begin{equation}
n_{\pi L} (E) = Z_T^2 \alpha {L [(2L+1)!!]^2 \over (2\pi)^3 (L+1)}  \sum_M |G_{\pi LM} (c/v) |^2  g_M (E),
\end{equation}
where 
\begin{equation}
g_M (E) = 2\pi \left({E  \over \gamma \hbar v}\right)^2 \int_0^\infty db \, b K_M^2\left({E  b\over \gamma \hbar v}\right) \exp[-2\chi_I(b)], \label{gme}
\end{equation}
with $\chi_I$ being the imaginary part of the eikonal phase in Eq. \eqref{eik7}.

A common misconception found in the literature is to assume that in a ``pure'' Coulomb excitation the angular distributions should not display a diffraction pattern because no nuclear interaction is considered. This is not true in quantum mechanics because even without nuclear excitation the scattering waves are still modified by absorption at small impact parameters. The angular distribution  depends on the integral in Eq. \eqref{Omega} which induces diffraction patterns in the angular distribution. The misconception arises because in the  frequently used semiclassical model the quantum scattering is replaced by classical trajectories and a smooth angular distribution emerges. I will show how the angular distribution in the semiclassical picture emerges from the quantum mechanical method described above, but only when the Coulomb interaction is overwhelmingly dominant.

\subsection{Semiclassical approximation at large impact parameters}
Using the same approximations as in Ref. \cite{BERTULANI1993158}, I can show that when $\chi_I=0$, i.e., $\chi=\chi_c$,  the following result holds at large impact parameters,
\begin{eqnarray}
\left| \Omega_M \right|^2 = {1\over k^2} \left({d \sigma \over d \Omega}\right)_{Ruth} K_M \left( {\omega b_0\over \gamma v}\right)
 \label{cceq20}
\end{eqnarray}
where
 \begin{eqnarray}
b_0&=&{Z_PZ_Te^2\over 2 E_{cm}\sin(\theta/2)}. 
\label{b0class}
\end{eqnarray}
Inserting this approximation into Eq. \eqref{virtpiL} yields smooth angular distributions for Coulomb excitation. The total cross sections are also simplified because the integral in Eq.   \eqref{gme}  can be done analytically if one restricts to a minimum impact parameter above which the Coulomb interaction is dominant \cite{BERTULANI1988299}.

So far, I did not discuss the form of the Coulomb and nuclear interactions entering Eq. \eqref{cceq4} as well as the model to calculate the projectile transition matrix elements.  I show this in the next Section, with emphasis on the Lorentz transformation properties.

\section{Relativistic Coulomb interaction}
Because the continuum wave functions extend to large distances, the continuum-continuum transition matrix elements may diverge if the Coulomb potential is not properly regularized.  
The covariant Coulomb interaction Hamiltonian for the breakup of a projectile is given by 
\begin{equation}
{\cal H}_{int}^{C}= j_\nu A^\nu = \sum_i \left[ q_i \Phi({\bf r}_i) - {1\over c} {\bf j}_i \cdot {\bf A}({\bf r}_i)\right], \label{amujmu}
\end{equation} 
where the index $i$ denotes the individual charges. $\Phi({\bf r}_i) $ and ${\bf A}({\bf r}_i)$ are the electric and vector potentials due to the target at the location of the charges within the projectile.  ${\bf j}_i $ is the charge current within the projectile. I include  convective currents only.  For a projectile moving at high speeds, the Coulomb deflection is small and one can resort to the approximation ${\bf A}({\bf r}_i)=({\bf v}/c)\Phi({\bf r}_i)$ where ${\bf v}$ is the projectile velocity. 
A proper description of ``close'' versus ``distant'' collisions for relativistic Coulomb excitation was developed in Ref. \cite{EsbensenPRC65.024605}.  I will consider electric excitations only and cast the Coulomb potentials in an adequate form. For E1 Coulomb excitations, correcting for the  center of mass motion of the projectile, the analytical form is
\begin{equation}
{\cal H}_{{\rm E1}\mu}={4\pi \gamma Z_Te_1e \over 3}   {r_< \over r^2_>}Y_{1\mu}\left(  \hat{\bf r}_< \right) Y^*_{1{\mu}}\left(  \hat{\bf r}_> \right), \label{eqE1}
\end{equation}
where $r_<$ ($r_>$) is the smallest (largest) value of $r$, the internal coordinate of the wave function of the projectile; the relativistic distance to the  target $r' = \sqrt{y^2 + \gamma^2z^2}$,  $e_1 = Z_a/m_a - Z_b/m_b$ is the effective charge for the projectile  with a clusterlike structure $A_P = a+b$; and $\gamma$ is the Lorentz factor. These potentials are amenable to a separation of {\it close} and {\it distant} collisions. In the former case the wave functions of the bound and continuum states extend beyond  the ``relativistic distance" between the projectile and target,  i.e.,  beyond $r' = \sqrt{y^2 + \gamma^2z^2}$.

For E2 excitations the Coulomb potential is given by
\begin{equation}
{\cal H}_{{\rm E2}\mu}={ 4\pi \gamma Z_Te_2e \over 5}  {r^2_< \over r^3_>}Y_{1\mu}\left(  \hat{\bf r}_< \right) Y^*_{1{\mu}}\left(  \hat{\bf r}_> \right), \label{eqE2}
\end{equation}
where $e_2=Z_a(m_b/m_P)^2+Z_b(m_a/m_P)^2$. 

The expressions above reduce to the usual Coulomb excitation expansion at low energies. The inclusion of the convective current, i.e., the second term in Eq. \eqref{amujmu}, is omitted in these equations. An exact account of the contributions from the convective current is presented in Ref. \cite{EsbensenPRC65.024605}. Convective currents are found to be negligible by a factor $(E_xr/\hbar v)^2$ for the cases discussed in this work, where $E_x$ is the excitation energy and $r$ is a distance from the projectile center of mass.  This conclusion also holds for continuum-continuum transitions for which $r$ can reach large values. Convergence in continuum-continuum transitions is guaranteed because the integrand in the matrix elements (discussed below) change the role of $r$ and $r'$ when one becomes larger than the other. The discretization of the continuum states according to the procedure described in Eqs. (\ref{gamm1}) and (\ref{gamm2}) also helps the quick convergence of the transition matrix elements.

 When inserted into Eq. \eqref{cceq4}, one will need the matrix elements $\left\langle \Phi_{c}|r^LY_{L\mu}(  \hat{\bf r})|\Phi_{c'}\right\rangle$ for distant collisions and  $\left\langle \Phi_{c}|Y^*_{L\mu}(  \hat{\bf r})/r^{L+1}|\Phi_{c'}\right\rangle$ for close collisions. In both cases, one can separate the geometric coefficients from the radial integrals by using
\begin{equation}
\left\langle \Phi_{c}|r^L Y_{L\mu}(  \hat{\bf r})|\Phi_{c'}\right\rangle = \left\langle J_cM_c L \mu  \big| J_{c'}M_{c'} \right\rangle  \left\langle \Phi_{c}||r^L Y_{L}||\Phi_{c'}\right\rangle , \label{coulmat}
\end{equation}
and similarly for the matrix elements of  $Y^*_{L\mu}(  \hat{\bf r})/r^{L+1}$.

For first-order excitations, I will only use the distant form of the Coulomb potential because the ground-state wave functions are well localized.  Using Eqs. (\ref{eqE1}) and (\ref{eqE2}) in Eq. \eqref{cceq9} it is straightforward to show that the Coulomb excitation cross sections are directly proportional to the electric dipole and quadrupole response functions, defined as
\begin{equation}
{dB(EL) \over dE_x} =  \frac{\left|\left\langle \Phi(E_c,J)||r^L Y_{L}||\Phi_{0}\right\rangle\right|^2}{2J_0+1} ,
\label{dbdei}
\end{equation}
with $E_x\equiv E_c-E_0$. The final state is labeled in terms of its energy $E_c$, and angular momentum  $J$, i.e., $\Phi(E_c,J)\equiv \Phi_c$. If the response function is not available from a theoretical model but can be extracted from an experiment, the matrix elements needed for the Coulomb excitation cross section can be approximated as 
\begin{equation}
\left\langle \Phi(E_c,J)|| r^L Y_{L}||\Phi_{0}\right\rangle = e^{i\delta} \left[ (2J_0+1)dB(EL)/ dE_x\right]^{1/2}, \label{redm1}
\end{equation}
where $\delta$ is a phase, which I conveniently choose to be equal to zero\footnote{The effects of the nuclear phases $\delta$ can be studied by generating random complex values and identifying their impact on the reaction observables.}. 
This procedure is also useful to obtain the excitation amplitudes for couplings between the ground and continuum states using the coupled channels equations Eq. \eqref{cceq4}.

The matrix elements for continuum-continuum coupling  cannot be extracted from an analysis of Coulomb breakup experiments. One has to rely on a theoretical model. For simplicity I will use continuum waves generated from a standard two-body Woods-Saxon potential. This is a departure from the three-body model adopted for the ground-state to continuum transition.
Using these two assumptions I assess the effects of ground-state-continuum and continuum-continuum  transitions, including all couplings between the states. It is worth mentioning that employing two-body continuum wave functions to obtain matrix elements for continuum-continuum transitions is inconsistent with the three-body wave functions used to calculate  ground-state to continuum transitions. This might be a source of uncertainties to account for the effects of continuum-continuum couplings. The sensitivity of the results on the parameters of the Woods-Saxon potential were found to be minor, except if the range of the potential is increased by about three times beyond typical nuclear potential ranges, which are close to the sum of the root-mean-square radii of the ground-state densities.

\begin{figure}[t]
\begin{center}
\includegraphics[
width=3.2in]
{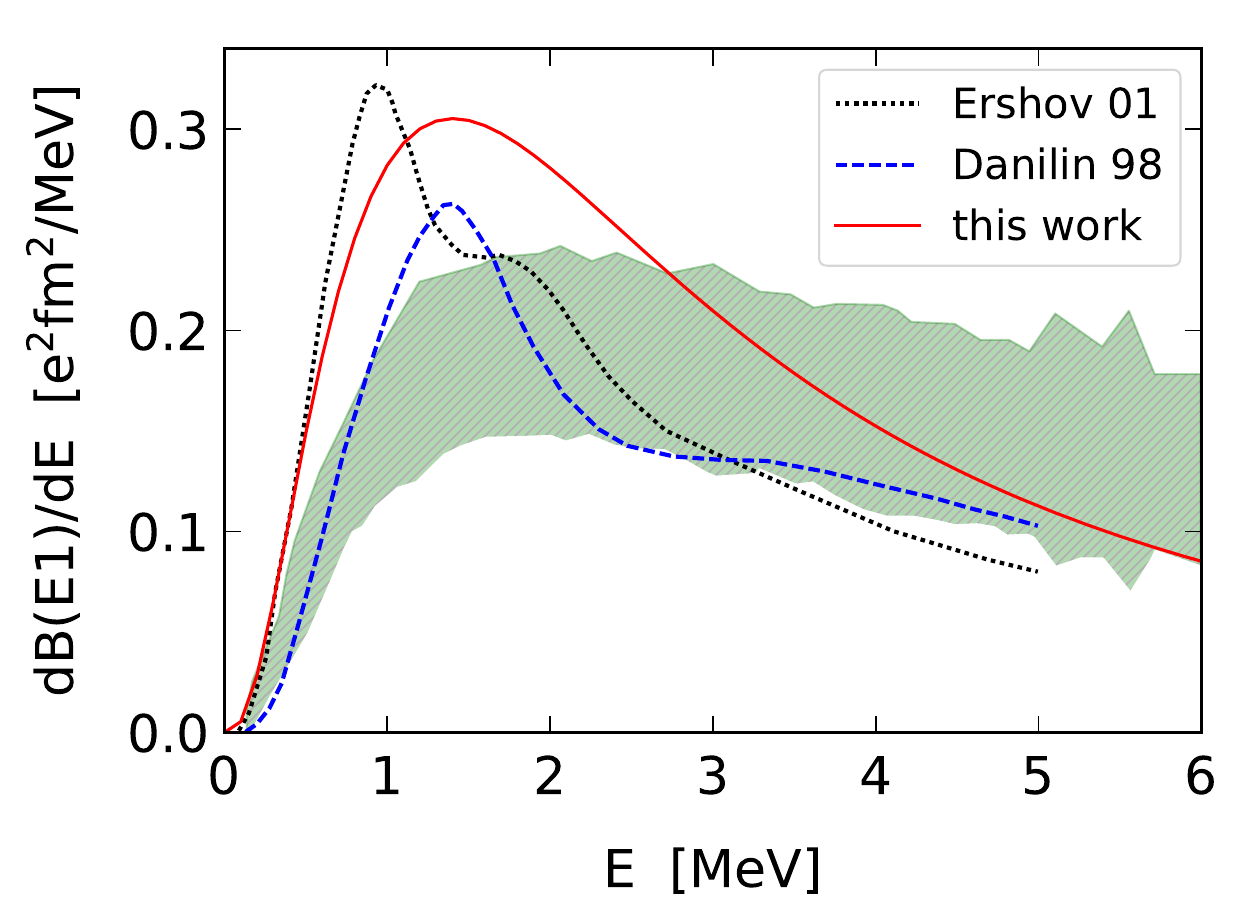}
\caption{Comparisons between theoretical predictions for the response function $dB(E1)/dE$, Eq. \eqref{dbdei}, with experimental data extracted from the Coulomb breakup of $^6$He projectiles incident on Sn and Pb targets  \cite{PhysRevC.59.1252}. Dashed and dotted curves are the results obtained with three-body calculations reported in Refs. \cite{DANILIN1998383,PhysRevC.64.064609,CobisPRL79.2411}. The model explained in the text is shown as a solid line. }
\label{dbde6he}
\end{center}
\end{figure}

\section{Relativistic nuclear interaction\label{rnd}}

\subsection{Real part of the nucleus-nucleus interaction}
At low energies ($E_{lab} \lesssim 50 $ MeV/nucleon), there are numerous methods used in the literature to describe optical potentials for nucleus-nucleus collisions. These potentials consist of a real and an imaginary part accounting for all processes leading to energy loss from the elastic scattering channel. However, the extension of this approach to direct nuclear reactions  is not well justified for relativistic collisions. Lorentz invariance implies that potentials should be four-vectors.  Moreover, due to retardation, a microscopic ab initio formalism aimed at building a  nucleus-nucleus potential from nucleon-nucleon interactions becomes cumbersome and unrealistic. A rather successful method, known as ``Dirac phenomenology,'' has been developed for nucleon-nucleus scattering \cite{ArnoldPRC.23.1949}. This phenomenological method captures the essence of the spin-orbit interaction in proton-nucleus scattering. It is based on the use of the Dirac equation with two nuclear potentials. The first is a potential $U_0(r)$ transforming as the timelike component of a Lorentz four-vector. The second potential,  $U_S(r)$, is  a Lorentz scalar. These potentials are viewed as effective interactions due to nucleon-nucleon scattering  via  meson exchange, folded with proton and neutron densities. The potentials depend on the masses and coupling constants of one-boson exchange of the neutral vector $\omega$ and scalar $\sigma$  mesons \cite{ArnoldPRC.23.1949}.

One could try to generalize this procedure from proton-nucleus to nucleus-nucleus scattering. However, additional complications arise because the projectile is now a composite object and the distance between projectile and target nucleons is affected by retardation. Unless an unjustified folding procedure is adopted, such a method is not better than adopting the same kind of phenomenology used for low energy nucleus-nucleus scattering.  The challenges for the formulation of retardation effects in the nucleus-nucleus potential was pointed out in Ref. \cite{PhysRevC.83.024907} using a relativistic mean field approach. This pioneer study has not been explored further. It has been shown in Ref.  \cite{PhysRevC.83.024907}  that, with the inclusion of retardation, the nucleus-nucleus potential is substantially modified at the nuclear surface. The nuclear surface is  the most relevant interaction region in direct nuclear reactions at high energies. 

 The usual formalism adopted for low energy reactions resorts to a nuclear potential for the projectile dissociation given by $U = \left( \sum_i U_{iT}\right)-U_{PT}$. This is a sum of the interactions between the core ($i=C$)  and the projectile clusters with the target $T$, minus the projectile-target interaction to eliminate the center-of-mass motion. In my present context, this amounts to find a relativistic optical potential $U_{iT}$, which is a daunting task.  Here I adopt a basic approach to the (real) nuclear potential that captures the essence of low-energy reactions and includes a modification due to relativity. At high energies the collision is peripheral in nature, due to absorption at small impact parameters. The nucleon-nucleon interactions are also short range. Therefore, I expect that only the tails of the nuclear potentials are effective and only the region around the nuclear surface of the projectile is substantially probed. This justifies adopting a Taylor expansion of the potential around the surface region of the projectile, a procedure similar to the one introduced in Ref. \cite{SATCHLER1987215}. For isovector excitations the potential is proportional to the neutron skin, $\Delta R_{np}$, namely ${\cal U} \sim \Delta R_{np} dU/dr$, where  $U$ is the nucleus-nucleus optical potential. The physics justification is that the strong interaction tends acts differently on protons and neutrons and tends to pull them apart. I propose without a rigorous proof that a similar dependence for the breakup potential arises due to its tendency to separate the core from the neutrons and the following equation  ensues for the nuclear breakup potential:
 \be
{\cal U} (r) =- \beta \left( \left<r^2_{He_C}\right>^{1/2}-\left<r^2_{He_{2n}}\right>^{1/2}\right)  {dU\over dr}, \label{nint}
\ee
where  $\left<r_{He_C}^2\right>^{1/2}$ ($\left<r_{He_{2n}}^2\right>^{1/2}$) is the root mean square  radius of the core (valence nucleons) in $^6$He. I have introduced a parameter $\beta$ to account for small deviations from the surface approximation. The averages above depend on the core and valence neutrons density  distribution in $^6$He, which will be discussed below.

The scalar nuclear potential $U$ entering Eq. \eqref{nint} is calculated with a double-folding approximation,
\begin{equation}
U({\bf r}) = \int \rho_P({\bf r}') v_{0} ({\bf s}) \rho_T({\bf r}'') d^3r' d^3 r'',
\end{equation}
where $v_0({\bf s})$ is the effective nucleon-nucleon potential, with  ${\bf s}={\bf r}'+{\bf r} -{\bf r}''$. 
For simplicity, I use the M3Y interaction \cite{BERTSCH1977399}, parametrized to reproduce inelastic nucleus-nucleus scattering at intermediate energies,
\ba
v_0(r) &=& 7999{\exp(-4r)\over 4r}-2134{\exp(-2.5r)\over 2.5r} \nonumber \\
&+&276\left(1-0.005{E\over A_P}\right)\delta(r),
\ea
in MeV and fm units. The potential ${\cal U}(r)$ is peaked at the surface of the projectile due to the derivative in Eq. \eqref{nint}.

To account for the effect of retardation, the nuclear interaction is modified so that $${\cal U}({\bf b},z) \rightarrow \gamma {\cal U}({\bf b},\gamma z),$$ mimicking the Lorentz effects in the Coulomb interaction. Although this prescription is not based on first principles, it captures again the essence of  nuclear-induced breakup at high energies. A variant of this procedure  has been used with success at low energies to describe collective excitations in nucleus-nucleus collisions \cite{SATCHLER1987215}.  

\subsection{Imaginary part of the nucleus-nucleus interaction}

The inclusion of the absorption channel due to an imaginary part of the interaction is taken into account with Eq. \eqref{eik7} and the profile function in Eq. \eqref{eik8}.  It is well known that the absorption caused by the  imaginary part of the optical potential can also lead to the breakup of the projectile, a process known as {\it diffraction dissociation}. The process leads to  low energy-momentum transfer above the threshold of fragmentation. If the momentum transfer is small enough ($q \ll 1/R$), where $R$ is the nuclear radius, the diffraction around the target couples coherently to nucleons and clusters inside the projectile.  The word {\it nuclear elastic} interaction, frequently found in the literature for this kind of process, is misleading because the real part of the potential can also induce the fragmentation of the projectile without changing the state of the target. This phenomenon is also part of the so-called nuclear elastic breakup. Diffraction dissociation is a purely wave-mechanical process caused by the diffraction of the projectile around an opaque matter distribution of the target, and is therefore an ``elastic'' process. I prefer to use the notation diffraction dissociation specifically for the breakup occurring due to the imaginary part of the nuclear potential, stemming from absorption at small impact parameters.

The diffraction dissociation formalism was introduced by Akhiezer and Sitenko to describe the dissociation of the deuterons by a ``black nucleus''  \cite{PhysRev.106.1236}. A similar formalism was also developed independently by Glauber \cite{PhysRev.99.1515}. An extension of the formalism to the dissociation of a generic weakly-bound nuclear projectile was formulated by Bertulani and Baur \cite{BBNPA1988480}.  It has been shown that a proper consideration of the binding energy of the projectile decreases considerably the magnitude of the diffraction dissociation cross section in contrast to the original estimates published in Refs. \cite{PhysRev.106.1236,PhysRev.99.1515}.

The diffraction dissociation of $^6$He for a momentum transfer ${\bf q}$ in a collision with the target depends on the ``survival probability" of the core, ${\cal S}_C$, and of the two neutrons, ${\cal S}_{2n}$, weighted with the probability amplitude to find the system initially at a distance ${\bf r}$,
\ba
{d \sigma_d\over d^3 {\bf q}} &=& {\eta \over (2\pi)^4} \int d^2 b_C \left| \int d^3 r {\cal S}_C(b_C){\cal S}_{2n}(b_{2n})\right. \nonumber \\
&\times& {e^{-\eta r} \over r}\left. \left[ e^{i{\bf q.r}}+ {1\over iq-\eta}{e^{-iqr}\over r}\right]\right|^2, \label{bb88}
\ea
where $b_{2n}$  can be written in terms of the impact parameter of the core $b_C$ and the relative distance between the $\alpha$-core and the two neutrons, ${\bf r} = (\boldsymbol{\rho},z)\equiv(r,\theta,\phi)$ in the form 
$$ b_{2n}= |{\boldsymbol \rho} -{\bf b}_{2n}|=\sqrt{r^2\sin^2\theta+b_C^2-2b_Cr\sin\theta \cos\phi}.$$ 
For the purposes of calculating the {\it S}-matrices for the neutrons+target, I assume that the two neutrons are treated as a di-neutron particle. 

The expression inside the brackets in Eq. \eqref{bb88} is the asymptotic wave function of the $\alpha$-2n system with $\eta = (2\mu S/\hbar^2)^{1/2}$, with $S$ being the separation energy. The second term inside the brackets account for the completeness of continuum and bound-state wave functions. $\mu=4m_N/3$ is the reduced mass of the $\alpha$ + 2n system. In contrast to the previous equations for {\it S}-matrices to account for absorption, as in Eq. \eqref{eik7}, the formalism described above requires  ${\cal S}_C$ for the core-target interaction and ${\cal S}_{2n}$ for the neutron-target interaction separately. The same Eq. \eqref{eik7} is used, but with $\rho_P$ replaced by $\rho_C$ and by $\rho_{2n}$, respectively.  For $\rho_C$ (alpha core in $^6$He) I use $\rho_C(r) = \rho_0\exp(-r^2/a^2)$, with $a =1.325$ fm, implying a rms radius $\sqrt{\left< r^2 \right>_\alpha}=1.62$ fm with $\rho_0$ adjusted so that $\int \rho_C(r) d^3r = 4$. The valence neutrons density is described by $\rho_{2n}(r) =\rho_0 r^2\exp(-r^2/b^2)$, with $b=2.045$ fm, implying a rms radius for the valence neutrons $\sqrt{\left< r^2 \right>_n}=3.23$ fm. These values are in accordance with the results presented in Ref. \cite{TANIHATA1992261} for the $^6$He density assumed to be a properly normalized sum of the core and neutron distributions  (see also Ref. \cite{PhysRevC.76.051602}). The neutron distribution is normalized to 2 before inserting in Eq.  \eqref{eik7} to calculate ${\cal S}_{2n}$.

In the sharp-cutoff approximation used in Ref. \cite{BBNPA1988480},  ${\cal S}_C{\cal S}_{2n}\simeq \Theta(r-R)$, where $R$ is the sum of the projectile and target rms radii, $R\sim \left<r^2\right>^{1/2}_{^6{\rm He}} + \left<r^2\right>^{1/2}_{^{208}{\rm Pb}}$. In this approximation, the integrals in Eq. \eqref{bb88} can be done analytically and the diffraction dissociation cross section for the excitation to a channel with energy $E$ is obtained as a sum of the independent scattering of the clusters by the target minus a term corresponding to the interference scattering of the clusters, also called the {\it eclipse}, or {\it shadowing}, term. The shadowing term tends to interfere destructively with the first two terms. 
For strongly bound nuclei,  $S/E \rightarrow \infty$, and one can show  \cite{BBNPA1988480} that  $d\sigma_d/dE \rightarrow 0$. On the other hand,  for loosely bound nuclei, $S/E \rightarrow 0$, and
\be
{d\sigma_{d}\over dE} \sim {2\pi R^2 \over E} J_1^2 \left({\sqrt{2\mu E}\over \hbar} R\right) ,\label{dsdelim}
\ee
meaning that the diffraction dissociation cross section in the very low biding regime is nearly equal to the sum of the elastic diffraction cross section of each cluster by the target. It also reveals the diffraction scattering pattern emerging from the Bessel function $J_1$. The angular distribution dips occur at multiples of $\theta \simeq 1/kR$, and  for excitation energies multiples of $E \sim \hbar^2/2\mu R^2$. Because diffraction dissociation only involves coordinates transverse to the incident projectile direction, it is also Lorentz invariant.  

It is important to mention that Eq. \eqref{dsdelim} is not applicable to the problem studied here.  In  the case of $^6$He, the nucleon separation energy is not negligible ($\approx 1$ MeV) compared to the dissociation energies $E$ and the full diffraction dissociation equation, Eq. \eqref{bb88}, is used with the {\it S}-matrices obtained from the folding of  nuclear densities, indicated in Eq. \eqref{eik7}. The diffraction dissociation is roughly  independent of the projectile incident energy for the range of energies considered in this work.

\section{Halo effective theory for the response function of $^6$He \label{HET}}

The purpose of this work is to report a study of reaction mechanisms in extracting the electromagnetic response of halo nuclei, such as $^6$He, in reactions at intermediate and high bombarding energies. The formalism described in the previous sections requires knowledge of matrix elements for the various transitions. I use a simplified three-body model for $^6$He as a n+n+$\alpha$ system based on halo effective theory (HET)\footnote{Not to be confused with halo Effective Field Theory, or halo-EFT, which uses concepts of field theory \cite{BertHamKol02}. HET uses the traditional  Schr\"odinger mechanics.}. In this model, the bound--state wave function in the center-of-mass system is written as an expansion over hyperspherical harmonics (HH) (see, e.g., Refs. \cite{PhysRevC.64.064609,CobisPRL79.2411,ZHUKOV1993151,A_Pushkin_1996,DANILIN1998383,FORSSEN2002639,FORSSEN200248,Kievsky_2008}), 
\begin{equation}
\Psi\left(  \mathbf{x},\mathbf{y}\right)  =\frac{1}{\rho^{5/2}}\sum_{KLSl_{x}l_{y}}\Phi_{KLS}^{l_{x}l_{y}}\left(  \rho\right)  \left[
\mathcal{J}_{KL}^{l_{x}l_{y}}\left(  \Omega_{5}\right)  \otimes\chi_{S}\right]  _{JM}. \label{hhwf}%
\end{equation}
In this equation, $\mathbf{x}$ and $\mathbf{y}$ are the Jacobi coordinate vectors  $\mathbf{x}=\frac{1}{\sqrt{2}}\left(\mathbf{r}_{1}-\mathbf{r}_{2}\right)$ and ${\bf y}=\sqrt{\frac{2\left(A-2\right)  }{A}}\left(\frac{\mathbf{r}_{1}+\mathbf{r}_{2}}{2}-\mathbf{r}_{c}\right)$, where A is the nuclear mass, $\mathbf{r}_{1}$ and $\mathbf{r}_{2}$ are the positions of the valence neutrons, and $\mathbf{r}_{c}$ is the position of the core. The hyperradius \ $\rho$ determines the size of a three body state: $\rho ^{2}=x^{2}+y^{2}$. The five angles $\left\{ \Omega_{5}\right\}  $ include usual angles $(\theta_{x},\phi_{x})$, $(\theta_{y},\phi_{y})$ which parametrize the direction of the unit vectors $\widehat{\mathbf{x}}$ and $\widehat{\mathbf{y}}$ and the hyperangle $\theta$, related by $x=\rho \sin\theta$ and $y=\rho\cos\theta$, where $0\leq\theta\leq\pi/2$.

The insertion of the three-body wave function, Eq. \eqref{hhwf}, into the Schr\"odinger equation yields a set of coupled differential equations for the hyperradial wave function
$\Phi_{KLS}^{l_{x}l_{y}}\left(  \rho\right) $. Assuming that the nuclear potentials between the three particles are known, this method delivers the bound-state wave function for a three-body system with angular momentum $J$. To simplify calculations, I will follow here a simpler HET procedure using the asymptotic part of the bound-state wave function and a  set of final states which include the proper coordinate space and energy dependence. For weakly-bound systems (the two-neutron separation energy in $^6$He is 0.975 MeV)  the hyperradial functions entering the expansion \eqref{hhwf} behave asymptotically as $ \Phi_{a}\left(  \rho\right) \longrightarrow\text{const.}\times \exp\left( -\eta \rho\right)$ when $\rho\longrightarrow \infty$, where the two-neutron separation energy is related to $\eta$ by $S_{2n}=\hbar^{2}\eta^{2}/\left(  2m_{N}\right)  $. This wave function has similarities with the two-body case, if $\rho$ is interpreted as the distance $r$ between the core and the two nucleons, treated as one single particle. But  the mass $m_{N}$ would have to be replaced by $2m_{N}$ if a simple two-body  model were used for $^6$He. A full three-body model is superior in accuracy because it includes interactions between the three particles without further approximations.  But due to the uncertainty in the two- and three-body potentials as well as the Pauli-blocking procedure used in the calculations, the effort does not justify the benefits. The HET model used here, based on the asymptotic behavior of the three-body wave function includes the main features of the three-body phase space and is enough for my purposes. 

Because only the core carries charge, in a three-body model,  the E1 transition operator is given by  ${\cal O} \propto yY_{1M} (\hat{\bf y})$. 
The $E1$ transition matrix element is obtained by a sandwich of this operator between $\Phi_{a}\left(  \rho\right)/\rho^{5/2}$ and scattering wave functions. I will use distorted scattering states, and the expression for the radial matrix element is
\begin{equation}
\mathcal{M}\left(  E1\right)  =\int dxdy\frac{\Phi_{a}\left(  \rho\right)
}{\rho^{5/2}}\ y^{2}x u_{p}\left(  y\right)  u_{q}\left(  x\right)  ,
\label{ie1}%
\end{equation}
where $u_{p}\left(  y\right)  =j_{1}\left(  py\right)  \cos\delta_{nc}-n_{1}\left(  py\right)  \cos\delta_{nc}$ is the core-neutron asymptotic continuum wave function, assumed to be a $p$-wave, and $u_{q}\left(  x\right) =j_{0}\left(  qx\right)\cos\delta_{nn}-n_{0}\left(  qx\right)  \cos \delta_{nn}$ is the neutron-neutron asymptotic continuum wave function, assumed to be an $s$-wave. The relative momenta are given by $\mathbf{q}=\frac{1}{\sqrt{2}}\left(\mathbf{q}_{1}-\mathbf{q}_{2}\right)$, and $\mathbf{p}=\sqrt{\frac{2\left( A-2\right)  }{A} }\left(\frac{\mathbf{k}_{1}+\mathbf{k}_{2}}{2}-\mathbf{k}_{c}\right)$.

The $E1$ strength function is proportional to the square of the matrix element in eq. \eqref{ie1} integrated over all momentum variables, except for the total continuum energy $E_{rel}=\hbar^{2}\left(  q^{2}+p^{2}\right)  /2m_{N}$. This procedure gives%
\begin{equation}
\frac{dB\left(  E1\right)  }{dE}={\cal C}\int\left\vert
\mathcal{M}\left(  E1\right)  \right\vert ^{2}E^{2}\cos^{2}\Theta\sin
^{2}\Theta d\Theta d\Omega_{q}d\Omega_{p}, \label{dbde3b}%
\end{equation}
where $\Theta=\tan^{-1}\left(  q/p\right)  $. 

The $^{1}$S$_{0}$ phase shift in neutron-neutron scattering is remarkably well reproduced up to center of mass energy of order of 5 MeV by the first two terms in the effective-range expansion $k\cot\delta_{nn}\simeq-1/a_{nn}+r_{nn}k^{2}/2.$ Experimentally these parameters are determined to be $a_{nn}=-18.6$ fm and $r_{nn}=2.7$ fm \cite{G_rdestig_2009,Machleidt_2001,PhysRevC.73.014002}.  The dominant p$_{3/2}$-wave scattering in the n-$^{4}$He ($^{5}$He) system shows a resonance at low energies \cite{ENDF}. I assume that this phase-shift can be described by the resonance relation 
$\sin\delta_{nc}=({\Gamma/2})/{\sqrt{\left(  E_{r}-E_{R}\right)  ^{2}+\Gamma^{2}/4}}$, with $E_{R}=0.8$ MeV and $\Gamma=0.65$ MeV \cite{ENDF,BertHamKol02}.  Most integrals in Eqs. \eqref{ie1} and \eqref{dbde3b} can be done analytically (see, e.g., \cite{FORSSEN2002639,FORSSEN200248}, leaving two remaining integrals which can only be performed numerically. 

The result of the calculation is shown  as a solid curve in Fig. \ref{dbde6he}.  In the same figure I show a comparison with the experimental data from Ref. \cite{Lehr}. Dashed and dotted curves are the results obtained with three-body calculations reported in Refs. \cite{DANILIN1998383,PhysRevC.64.064609,CobisPRL79.2411}. The solid line is the response calculated with the HET model described above.
It is clear that the models described in Refs. \cite{DANILIN1998383,PhysRevC.64.064609,CobisPRL79.2411} do not reproduce the experimental response function in a reasonable way. The HET model described above does not do much better, but it roughly peaks around the same region as the data and has a similar shape. I have adjusted the normalization constant in Eq. \eqref{dbde3b} to best fit the experimental data.

\begin{figure}[t]
\begin{center}
\includegraphics[
width=3.5in]
{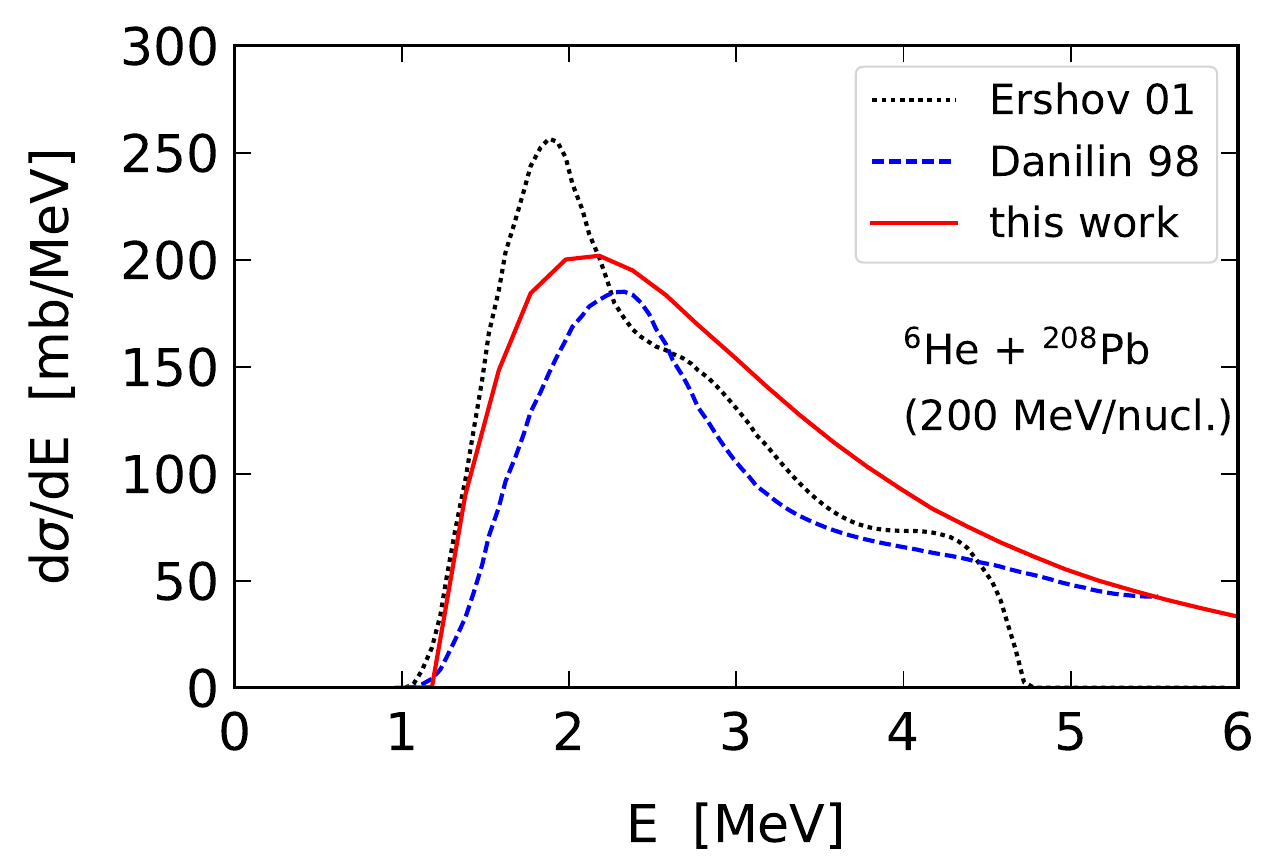}
\caption{Coulomb breakup cross section, $d\sigma/dE$, of $^6$He projectiles incident on Pb targets \cite{Lehr}. Also shown by means of dashed and dotted curves are the theoretical results for first-order Coulomb breakup using the response functions obtained with three-body calculations \cite{DANILIN1998383,PhysRevC.64.064609,CobisPRL79.2411} presented in Figure \ref{dbde6he}. }
\label{dsdePb}
\end{center}
\end{figure}

\section{Application to the breakup of $^6$He\label{bu6He}}

\subsection{Effects of channel coupling and nuclear breakup}

Coulomb excitation to first-order, with Eq. \eqref{dbdei} to determine the reduced matrix elements, yields the same results as the virtual photon method described in Ref. \cite{BERTULANI1988299}. If channel coupling is relevant, the phases of the reduced matrix elements in Eq.  \eqref{redm1} should be considered.  But, as I will show later, first-order calculations account for the bulk of the Coulomb breakup cross sections. Therefore, precise values of phases in the reduced matrix elements of Eq.  \eqref{redm1} should not be of major importance for the breakup of $^6$He.

\begin{figure}[tbh]
\begin{center}
\includegraphics[
width=3.4in]
{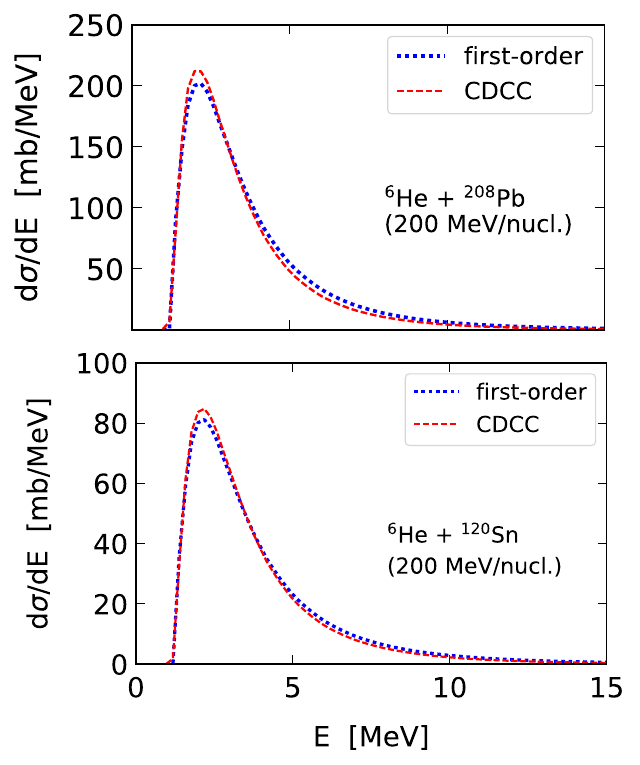}
\caption{Comparison between first-order and coupled-channels calculations of the Coulomb breakup cross sections, $d\sigma/dE$, of $^6$He projectiles incident on Pb (upper panel) and Sn (lower panel) targets \cite{Lehr}.  }
\label{cc}
\end{center}
\end{figure}

\begin{table}[h]
\begin{tabular}{|c|c|c|c|c|c|c|c|c|c|c|}
\tableline\tableline
Reaction & $E_{lab}$& $\sigma^{1st}_C$ &$\sigma^{cc}_C $ &$\sigma_{nuc}$&$\sigma_{dd}$ \\
& [MeV/nuc]  & [mb] & [mb]  &[mb] & [mb] \\
\hline\tableline
$^6$He + $^{208}$Pb & 200&635.79&	661.97& 36.76&26.59\\
& 1000& 410.71& 412.64&53.07&16.59\\
$^6$He + $^{120}$Sn & 200&269.53	&274.55&23.81&15.25	\\
& 1000& 164.72&165.82&48.24&11.25 \\
$^6$He + $^{12}$C & 200&5.54	&5.54&12.46&8.96	\\
& 1000& 3.05&3.05&28.13&8.96 \\
\tableline\tableline
\end{tabular}
\caption{Total Coulomb breakup cross sections for $^6$He + $^{208}$Pb,  $^6$He + $^{120}$Sn and $^6$He + $^{12}$C.  First-order Coulomb dissociation cross sections are denoted by  $\sigma^{1st}_C$ and CDCC calculations  are denoted by $\sigma^{cc}_C$. Breakup due to the real part of the nuclear interaction and due to diffraction dissociation  are labeled by $\sigma_{nuc}$ and $\sigma_{dd}$, respectively. \label{table1}}
\end{table}

\begin{figure}[tbh]
\begin{center}
\includegraphics[
width=3.4in]
{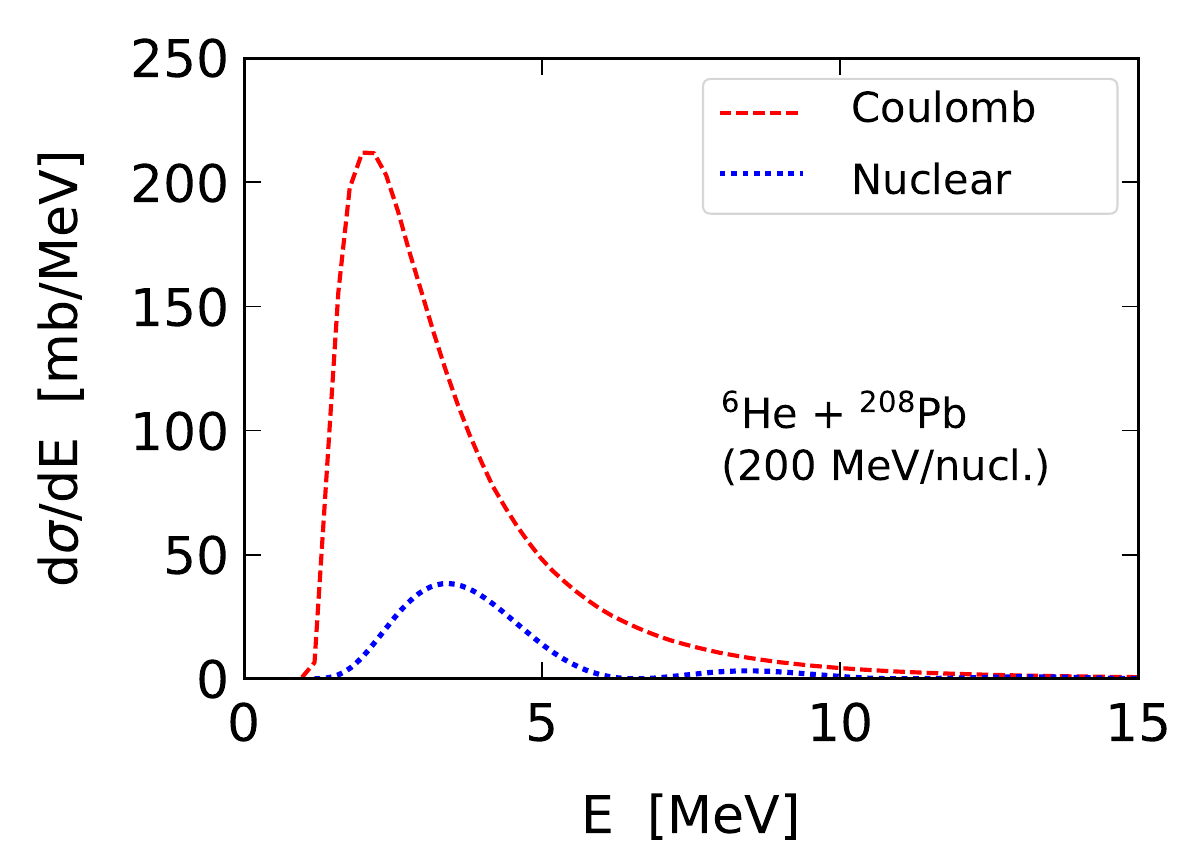}
\caption{Comparison between Coulomb breakup (dashed-line) of $^6$He projectiles incident on Pb targets at 200 MeV/nucleon with the nuclear dissociation, including diffraction dissociation (dotted line).}
\label{cnint}
\end{center}
\end{figure}

\begin{figure}[t]
\begin{center}
\includegraphics[
width=3.2in]
{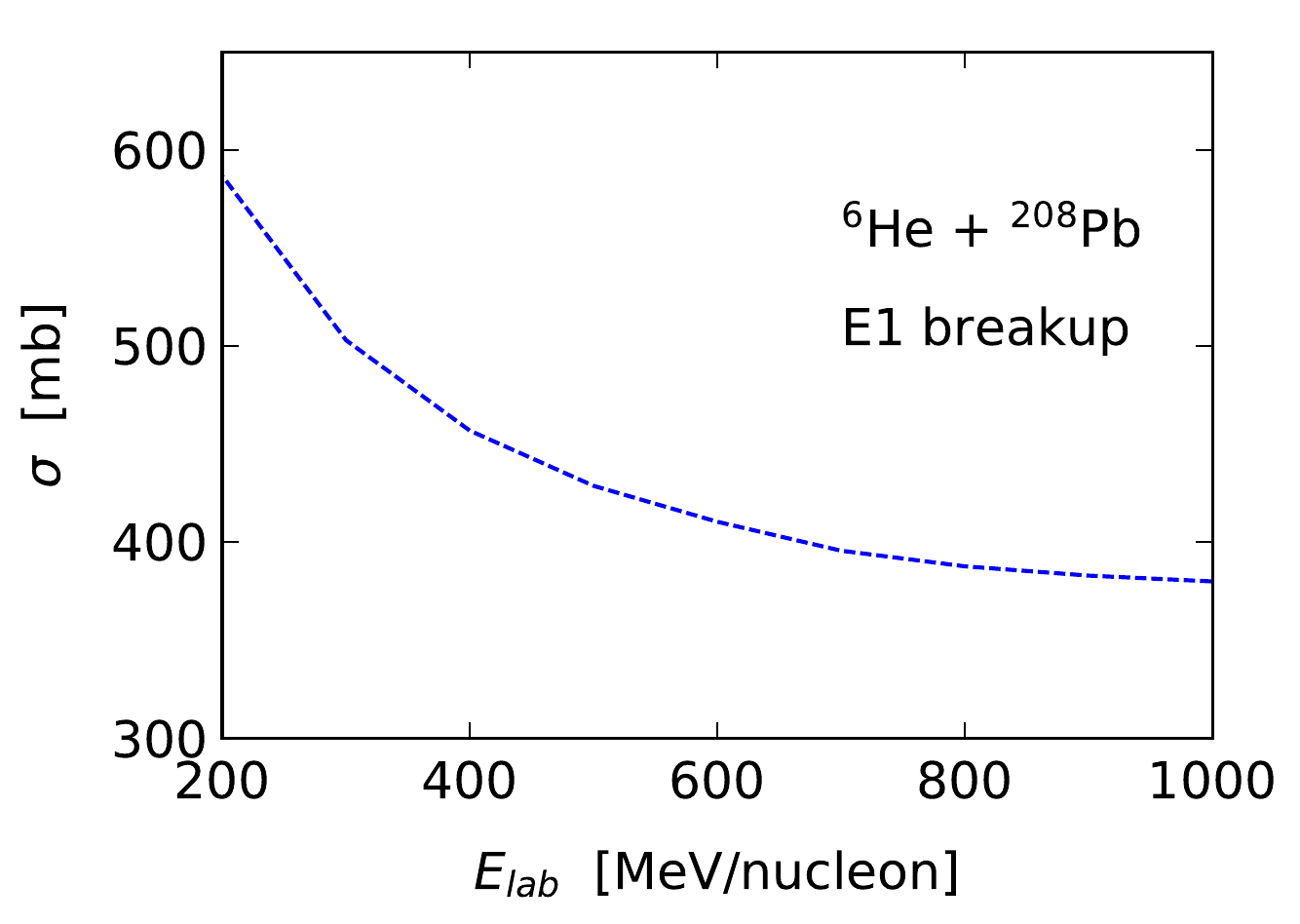}
\caption{Coulomb breakup cross sections of $^6$He projectiles incident on Pb targets as a function of the laboratory bombarding energy.}
\label{E1bup}
\end{center}
\end{figure}

To assess the impact of channel coupling including transitions in the continuum I will use the HET model described in Sec. \ref{HET} for the  response functions. This is shown in Fig. \ref{cc}  for the Coulomb breakup cross section, $d\sigma/dE$, of $^6$He projectiles incident on Pb and Sn targets \cite{Lehr}. One observes a small modification of the cross section around its peak values due to stronger transitions to states with large response. The first-order calculations are very close to RCDCC calculations, especially for low-$Z$ targets. The small enhancement of the differential cross sections at small energies is partially compensated by a small increase at larger energies. In Table \ref{table1} I show the different contributions of Coulomb, nuclear, and diffraction dissociation cross sections for $^6$He + $^{208}$Pb,  $^6$He + $^{120}$Sn and $^6$He + $^{12}$C at  200 and 1000 MeV/nucleon. I separate Coulomb from nuclear excitations by switching off in the calculations either the nuclear potential of Eq. \eqref{nint} or setting to zero the Coulomb matrix elements in Eq. \eqref{coulmat}. The diffraction dissociation cross section is calculated separately, using Eq. \eqref{bb88}.

As shown in Table \ref{table1} the inclusion of channel-coupling slightly increases ($\lesssim 4\%$) the total cross sections, mainly due to second-order transitions around the peak region, where the strength is concentrated.  It also shows that the Coulomb dissociation cross sections decrease with increasing bombarding energy, at least in the energy interval considered here. The physics reason is that as the bombarding energy increases, more (virtual) photons with energy higher than 2 MeV and fewer photons at lower energies, become available. The response function (Fig. \ref{dbde6he})  is smaller at large energies thus explaining the reduction of the cross section.  

Figure \ref{cnint} shows a comparison between Coulomb breakup (dashed-line) of $^6$He projectiles incident on Pb targets at 200 MeV/nucleon with the nuclear dissociation (dotted line) including real and imaginary parts (diffraction dissociation).   It is evident that the nuclear  contribution to the breakup is smaller ($\lesssim 10\%$ of the total cross section) in the region where the E1 response is of relevance. Also Coulomb-nuclear interference is found to be much smaller (by a factor $10^{-4}$) than both Coulomb and nuclear cross sections in the energy range explored here.  I have also calculated the electric quadrupole (E2) response $dB(E2)/dE$ and the corresponding Coulomb breakup cross sections, $d\sigma_C^{E2}/dE$, using the three-body model described above. I found that the cross sections for the E2 breakup model  are a factor $10^4$ smaller than the corresponding $E1$ breakup and can therefore be ignored.

One of the main difficulties in using breakup reactions to extract the response functions of radioactive projectiles lies in the fact that the corrections due to the nuclear interaction are not well known. My discussion in Sec. \ref{rnd} clearly highlights the difficulties in handling the nuclear interaction in high energy collisions. One commonly uses the strategy to scale the Coulomb breakup cross sections with the square of the target charge $Z_T^2$ and use a light target such as carbon to eliminate, at least partially, the corrections due to the breakup induced by the nuclear interaction. Our calculations displayed in Table \ref{table1} indicate that at 200 MeV/nucleon the ratios of the cross sections  are  $\sigma_{Pb}/\sigma_{Sn}=2.36$ and $\sigma_{Sn}/\sigma_{C}=48.6$ whereas $Z_{Pb}^2/Z_{Sn}^2 = 2.67$ and $Z_{Sn}^2/Z_{C}^2 = 69.4$, respectively. At 1000 MeV/nucleon, I get $\sigma_{Pb}/\sigma_{Sn}=2.49$ and $\sigma_{Sn}/\sigma_{C}=50.0$. This points to a non-negligible dependence of the cross sections on the geometry of the reacting nuclei. 

In Fig. \ref{E1bup} I show the E1 Coulomb breakup cross section of  $^6$He projectiles  as a function of the bombarding energy in the range 200-1000 MeV/nucleon. One sees that at 200 MeV/nucleon the cross section is largest. This seems to be the ideal bombarding energy region for experimental measurements,  a fact explored in a recent experiment performed at the GSI, Germany \cite{Lehr}.

\begin{figure}[t]
\begin{center}
\includegraphics[
width=3.in]
{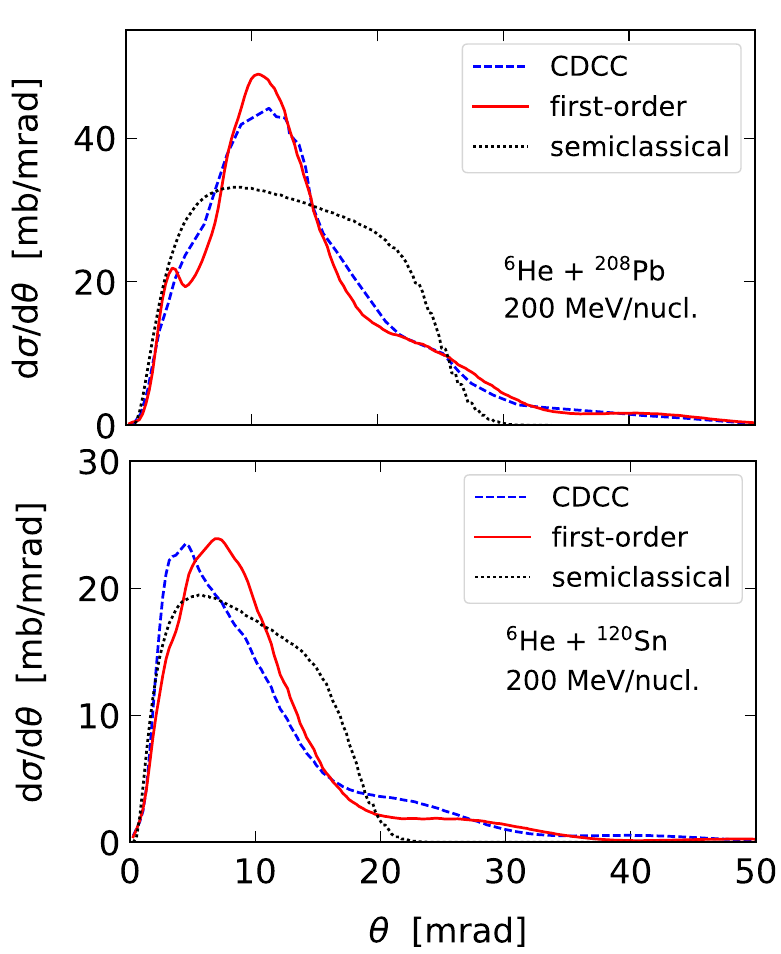}
\caption{Comparison between first-order and coupled-channels calculations for the Coulomb breakup angular distribution, $d\sigma/d\theta$, of $^6$He projectiles incident on Pb (upper panel) and Sn (lower panel) targets at 200 MeV/nucleon. The dashed line is a RCDCC calculation including continuum-continuum couplings.  It displays a diffraction pattern due to the nuclear absorption at small impact parameters. The solid (red) line is the result of first-order transitions only, using Eqs.  (\ref{cceq9}-\ref{cceq7}).  The dotted line is a semiclassical calculation,  based on Eqs.   (\ref{virtpiL},\ref{cceq20},\ref{b0class}).   }
\label{angdis}
\end{center}
\end{figure}

\subsection{Angular distributions}

I have checked how the angular distributions are affected by the different treatments of Coulomb scattering described in Section \ref{QScat}. In Figure \ref{angdis} I show a comparison between first-order and coupled-channels calculations for the Coulomb breakup angular distribution, $d\sigma/\theta$, of $^6$He projectiles incident on Pb (upper panel) and Sn (lower panel) targets at 200 MeV/nucleon. The dashed line is an RCDCC calculation including continuum-continuum coupling, using the formalism developed in Section \ref{cceqs}.  The RCDCC calculations  display a diffraction pattern due the nuclear absorption at small impact parameters. The solid (red) line is the result of first-order transitions only, using Eqs. (\ref{d2sfirst}-\ref{gme}).  One sees that calculations in first-order perturbation theory are also affected by a diffraction pattern due to absorption.  The dotted line is a semiclassical calculation, based on Eqs.   (\ref{virtpiL},\ref{cceq20},\ref{b0class}).  

The semiclassical angular distribution is smooth, starting from zero due to the inability of the Coulomb field to dissociate the projectiles in collisions at large impact parameters and dropping to zero again at large angles due to the absorption at small impact parameters.  The maximum occurs around an angle dictated by the adiabaticy parameter being close to the unity, i.e., when $\xi = Eb_0/ \gamma \hbar v\sim 1$ [$b_0$ is defined in Eq. \eqref{b0class}]. When $\xi$ is much larger than unity (small angles), the dynamic Coulomb field is not strong enough to breakup the projectile. On the other hand, when $\xi$ is much smaller than 1 (larger angles), absorption sets in \cite{BERTULANI1988299}. I also notice that the angular distributions are mildly sensitive to higher-order couplings. The integrated cross sections remain nearly unchanged, in accordance with my previous findings (see Fig. \ref{cc}). Significant modifications due to higher-order couplings are seen around the maximum of the angular distribution consistent with expectations. I also notice that the simple semiclassical method is not appropriate to describe the angular distributions. 

\section{Conclusions \label{concl}}

Relativistic Coulomb excitation of fast projectiles has long been a useful tool to unveil the properties of rare nuclear isotopes with applications to nuclear astrophysics. In this work I have studied the contributions of higher-order continuum-continuum couplings to the breakup of $^6$He projectiles in the bombarding energy range of 200-1000 MeV/nucleon. These effects were found to be small. In general, I found that the best energy regime to extract the electric dipole (E1) response of $^6$He is around 200 MeV/nucleon. 

More critical is the contribution of the nuclear interaction to the breakup. I have shown that this is a non-trivial task if the effects of retardation are of relevance. And in  fact they are because the nuclear mass increases by  20-100\%  in the bombarding energy regime studied here. The relativistic effects of the strong nucleus-nucleus interaction are manifest not only in the relativistic kinematics used in experimental analysis, but also in the relativistic dynamics used in the theoretical framework to analyze the data. At present, no widely accepted theory exists to treat this often ignored problem. I have shown that one can include some ingredients of relativity by small modifications in the traditional non-relativistic methods. Despite these issues being of relevance in Coulomb and nuclear excitation at intermediate energies, the nuclear breakup contributes $\lesssim 10\%$  to the cross sections involving $^6$He projectiles at 200 MeV/nucleon.

I have also shown that the angular distribution of the center of mass of the $^6$He fragments is slightly modified by the inclusion of higher-order terms. But one has to include absorption properly, otherwise it does not reflect the diffraction patterns characteristic of angular distributions. The total breakup cross section remains approximately unchanged from the one obtained in first-order perturbation theory. This is good news because first-order perturbation is much easier to handle than coupled-channels calculations.  CDCC calculations are also strongly dependent on the theoretical model adopted for the transition matrix elements. Here I have used simplifying models, to achieve practical results.  Finally, I have proved that semiclassical methods, frequently included in experimental analysis, are not appropriate to study angular distribution of the fragments.

\begin{acknowledgements}
The author has benefited from useful discussions with Thomas Aumann. He acknowledges support by the U.S. DOE Grant No. DE-FG02-08ER41533 and the Helmholtz Research Academy Hesse for FAIR.
\end{acknowledgements}

%\bibliographystyle{unsrt}
%\bibliography{Bertulani}

\end{document}